\documentclass[]{aastex631}

\def\tab{Table~}
\def\fig{Fig.~}

\def\sun{$_\odot$}
\def\um{$\rm \mu m$}
\def\h2{H$_2$}
\begin{document}

\title{Cloud-cloud collision and cluster formation in the W5-NW complex}

\author[0000-0002-7881-689X]{Namitha Issac}\thanks{E-mail:namiann@gmail.com}
\affiliation{Shanghai Astronomical Observatory, Chinese Academy of Sciences, 80 Nandan Road, Shanghai 200030, People's Republic of China}

\author{Anindya Saha}
\affiliation{Indian Institute of Space Science and Technology, Thiruvananthapuram 695 547, Kerala, India}

%\collaboration{20}{(AAS Journals Data Editors)}

\author{Saanika Choudhary}
\affiliation{Department of Physics and Astronomy, Northwestern University, Evanston, Illinois, 60208, USA}

\author{Aakash Chaudhary}
\affiliation{Indian Institute of Space Science and Technology, Thiruvananthapuram 695 547, Kerala, India}

\author[0000-0001-5917-5751]{Anandmayee Tej}\thanks{E-mail:tej@iist.ac.in}
\affiliation{Indian Institute of Space Science and Technology, Thiruvananthapuram 695 547, Kerala, India}

\author{Hong-Li Liu}
\affiliation{Department of Astronomy, Yunnan University, Kunming 650091, Peoples Republic of China}

\author{Tie Liu}
\affiliation{Shanghai Astronomical Observatory, Chinese Academy of Sciences, 80 Nandan Road, Shanghai 200030, People's Republic of China}

\author{Maheswar Gopinathan}
\affiliation{Indian Institute of Astrophysics, Koramangala II Block, Bangalore 560 034, India}



\begin{abstract}

We present a detailed structural and gas kinematic study of the star-forming complex W5-NW. A cloud-cloud collision scenario unravels with evidences of collision induced star and cluster formation. Various signatures of cloud-cloud collision such as ``complementary distribution'' and ``bridging-features'' are explored. At the colliding region, the two clouds have complementary morphologies, where W5-NWb has a filamentary key-like shape which fits into the U-shaped cavity in W5-NWa that behaves like a keyhole. The interaction region between the two clouds is characterised by bridging features with intermediate velocities connecting the two clouds. A skewed V-shaped bridging feature is also detected at the site of collision. A robust picture of the molecular gas distribution highlighting the bridges is seen in the position-position-velocity diagram obtained using the SCOUSEPY algorithm. Star cluster formation with an over-density of Class~I and Class~II young stellar objects is also seen towards this cloud complex, likely triggered by the cloud collision event. 

\end{abstract}



\section{Introduction} \label{sec:Introduction}
Star and cluster formation triggered by cloud-cloud collision (CCC) has been in focus in recent years \citep{2021PASJ...73S...1F}. Galactic scale numerical simulations show that cloud collisions are frequent events in gas-rich galaxies \citep{{2009ApJ...700..358T},{2015MNRAS.446.3608D}}.  
Theoretically, formation of dense cores, the star-forming `seeds', in the shock-compressed interface of colliding clouds has been seen in hydrodynamical simulation studies \citep{{1992PASJ...44..203H}, {2010MNRAS.405.1431A},{2014ApJ...792...63T}}. While such shock compressions lead to smaller Jeans mass and hence low-mass cores, high-mass star formation can be understood from magnetohydrodynamic (MHD) simulations. By introducing the effect of magnetic fields, \citet{2013ApJ...774L..31I} show that these supersonic collisions result in amplification of the magnetic field due to shock compression and enhanced turbulence in the interacting, compressed region. This goes in favour of massive filament formation \citep{2018PASJ...70S..53I} eventually fragmenting to dense, massive cores leading to the formation of high-mass stars. The number of cores created is a function of the relative collision velocity between the interacting clouds, whereas the growth of these cores is governed by how long they remain in the high-density shock compressed region \citep{2014ApJ...792...63T}. 
More recently, \citet{2022ApJ...940..106A} carried out shock compression simulations of molecular clouds using 3D isothermal MHD code with self-gravity. These authors demonstrated that long-shock-duration models result in the formation of massive and dense filaments, leading to the formation of high-mass stars. At peak column densities of the order of $\rm 10^{23}\,cm^{-2}$ several OB stars or massive star clusters can form at the shocked compressed layer undergoing gravitational collapse. Furthermore, MHD simulations by \citet{2023MNRAS.522.4972S} have also shown that magnetic field plays competing roles in the formation of massive cores, where they aid in mass accumulation during the collision process and hinder the growth after the collision. The duration of the collision essentially decides which effect of magnetic field plays the more crucial role. Thus, the longer the collision time, the more massive are the cores formed.

Recent observational studies provide ample evidence supporting the key role played by CCC in star formation. Studies have shown induced high-mass star and cluster formation \citep[e.g.][]{{2020MNRAS.499.3620I},{2021PASJ...73S...1F}}, formation of super star clusters \citep[e.g.][]{{2016ApJ...820...26F},{2014ApJ...780...36F}}, low- and intermediate-mass stars \citep[e.g.][]{{2017ApJ...835L..14G},{2019A&A...632A.115G}} and massive star clusters \citep[e.g.][]{{2018PASJ...70S..43S},{2020MNRAS.499.3620I}}. CCC resulting in the formation of single O-type stars with surrounding H\,{\small II} regions has also been observed \citep[e.g.][]{{2017ApJ...835..142T},{2015ApJ...806....7T}}. \citet{2021PASJ...73S..75E} for the first time explored the signatures of CCC in the Galactic Center by applying the methodology used to identify CCC in the Galactic disk. To compare the properties of CCC in the Galactic Center to the disk, they derived the physical parameters of the reported CCC candidates and established a relation between the relative velocity, peak column density, and the number of high-mass stars formed. They found that higher relative velocity would require higher column density to induce high-mass star formation, and the number of massive stars formed is also proportional to the peak column density (refer Fig. 9 of \citealt{2021PASJ...73S..75E}).

In this paper, we investigate the active star-forming region W5-NW (IRAS 02459+6029) which belongs to the well studied W3/W4/W5 molecular cloud complex \citep{{2008ApJ...688.1142K},{2012A&A...546A..74D}}. This complex is located at a distance of 2\,kpc \citep{{2006Sci...311...54X},{2006ApJ...645..337H}} in the Perseus spiral arm of the Milky Way.
In their extensive survey of W5 complex using data from \textit{Spitzer} telescope, \citet{2008ApJ...688.1142K} discussed the presence of at least two distinct generations of star formation and suggested triggering as a plausible mechanism to explain it. Further, these authors used photometry of a large number of point sources to analyze the spatial distributions of young stars, establish their evolutionary status and their clustering properties across this large star-forming region. In their investigation of W5-E using \textit{Herschel} data, \citet{2012A&A...546A..74D} aimed to understand the influence of role of H\,{\small II} regions on the star formation process in the vicinity. Based on their study, they have also indicated that triggered star formation is at work in this region.
From CO molecular line observations with the 13.7\,m Millimeter Telescope of the Purple Mountain Observatory (PMO) and the 3\,m K\"{o}lner Observatorium f\"ur Sub-Millimeter Astronomie (KOSMA) telescope, respectively, \citet{2008ChJAA...8..433X} and \citet{2012RAA....12.1269L} propose W5-NW to be a CCC region. 
These authors report the over density of YSOs in the possible collision region and also discuss that the two identified components in CO are adjacent both in space and velocity.
With major advancements in both observational and simulation studies \citep[][and references there in]{{2021PASJ...73S..75E},{2021PASJ...73S...1F}}, more distinct CCC identification methodology (e.g. complementary distribution) can be implemented. With this aim, we explore the W5-NW complex with the available higher resolution CO\,($3-2$) molecular line observations carried out with the {\it James Clerk Maxwell Telescope (JCMT)}. The CO\,($3-2$) transition with a higher critical density compared to CO\,($1-0$) will be a better tracer of relative high density regions, especially the shock compressed layer formed due to CCC \citep[e.g.][]{2017ApJ...835..142T}.
Such studies will not only contribute to the growing statistics of CCC events, but will also enable a comprehensive picture on the role of CCC and star formation in the Galaxy.

In presenting our analysis, the paper is structured in the following manner. In Section~\ref{sec:data} we give the details of the datasets used for this study. 
Results derived from these data are highlighted in Section~\ref{sec:results}. 
Section~\ref{sec:discussion} outlines the various signatures of CCC and how they manifest in the case of W5-NW and also discusses the possibility of collision induced cluster formation in the W5-NW star-forming complex. Section~\ref{sec:conclusion} summarizes the results and interpretations from this study.
 
\section{Observations}
\label{sec:data}
The analysis carried out in this study is based on archival datasets from {\it JCMT} and {\it Herschel Space Observatory} that are briefly described in this section.

\subsection{\textit{JCMT} CO data}
$^{12}$CO\,($3-2$) (Project ID: M07BH45B) and $^{13}$CO\,($3-2$) (Project ID: M09BU04) line data taken from the {\it JCMT} Science Archive are used to investigate the morphology and gas kinematics of the molecular cloud complex associated with the star-forming region W5-NW. These molecular line observations were carried out using the Heterodyne Array Receiver Program \citep[HARP;][]{2009MNRAS.399.1026B} on {\it JCMT} in scan mode. Operated by the East Asian Observatory, {\it JCMT} is a 15\,m telescope that functions in the  submillimeter wavelength regime. HARP is a Single Sideband array receiver. It can be tuned between 325 and 375 GHz and has an instantaneous bandwidth of $\sim$2 GHz and an Intermediate Frequency of 5\,GHz. It comprises of 16 receptors arranged in a 4$\times$4 grid, with an on-sky projected beam separation of 30 arcsec. At 345\,MHz, the main beam efficiency, $\eta_{\rm mb}$ is 0.64 and the beam size is 14$''$ \citep{2009MNRAS.399.1026B}. The $^{12}$CO\,($3-2$) and $^{13}$CO\,($3-2$) data cubes have spectral resolutions of 0.42 and 0.06\,km\,s$^{-1}$, and {\it rms} per channel of 1.1 and 6.5\,K, respectively.

\subsection{{\it Herschel} far-infrared maps}
We have used data from the {\it Herschel Space Observatory} archives to probe the cold dust emission associated with the W5-NW region.
This region 
was observed as a part of the {\it Herschel} Infrared Galactic Plane Survey \citep[Hi-Gal;][]{2010PASP..122..314M}, where the observations were carried out in parallel mode covering the wavelength range of 70 -- 500\,{\um}. The images in the Hi-Gal survey were obtained with the Photodetector Array Camera and Spectrometer \citep[PACS;][]{2010A&A...518L...2P} and Spectral and Photometric Imaging Receiver \citep[SPIRE;][]{2010A&A...518L...3G}. The retrieved images have resolutions of 5$''$, 13$''$, 18$''$, 24$''$, and 36$''$, and pixel sizes of 3$''$, 3$''$, 6$''$, 10$''$, and 14$''$
at 70, 160, 250, 350, and 500\,{\um}, respectively.
To generate the column density map (see Section~\ref{sec:signatures}), initial reduction steps were carried out using the Herschel Interactive Processing Environment (HIPE), which involves image unit conversion from MJy\,sr$^{-1}$ to Jy\,pixel$^{-1}$, followed by convolution and regridding of all the maps to the lowest resolution (36$''$) and pixel size (14$''$) of the 500\,{\um} image. The convolution kernels are taken from the \citet{2011PASP..123.1218A}.

\section{Results and Analyses}
\label{sec:results}

\subsection{CO line emission from W5-NW}
\label{sec:result1}
%%%%%%%%%%%%%%%%%%%%%%%%%%%%%%%%%%%%%%%%%
\begin{figure}
\centering 
\includegraphics[scale=0.5]{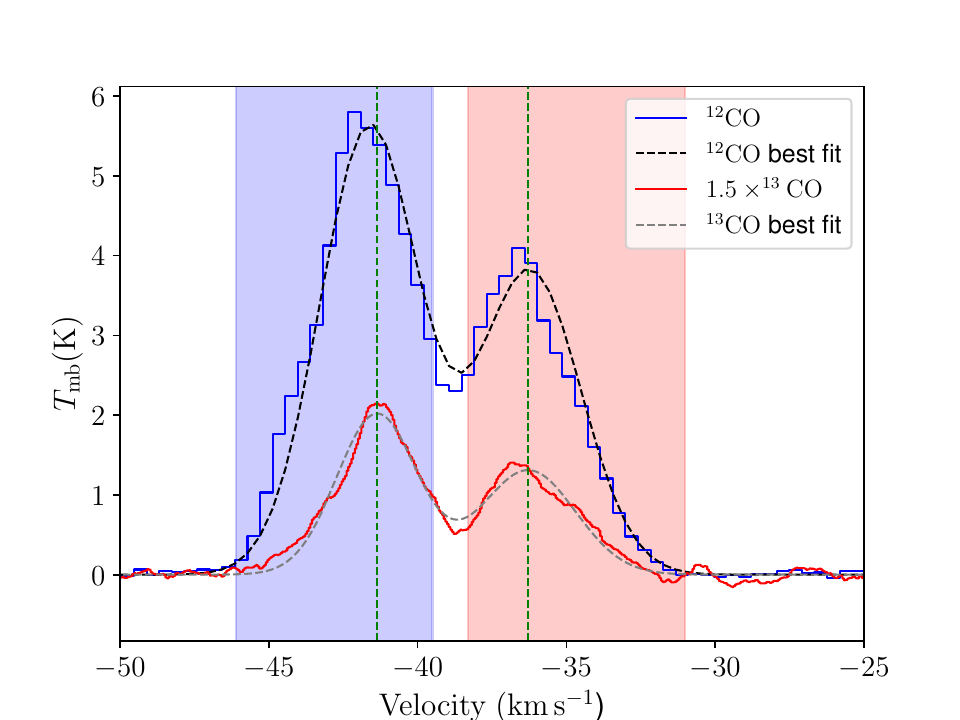} 
--\caption{Average spectra of $^{12}$CO\,($3-2$) (blue) and $^{13}$CO\,($3-2$) (red) towards W5-NW. The extracted spectra is averaged over the circular region shown in {\fig}\ref{fig:CD_temp} that covers the entire W5-NW complex. The $^{13}$CO\,($3-2$) spectrum is scaled-up by 1.5 times and is boxcar smoothed by 20 channels that corresponds to a velocity smoothing of $\sim 1.1\,\rm km\,s^{-1}$. The double Gaussian fit to the $^{12}$CO\,($3-2$) and $^{13}$CO\,($3-2$) lines are depicted by dashed black and grey curves, respectively. The blue and red shaded regions highlight the velocity ranges over which the integrated intensity maps are constructed for W5-NWa and W5-NWb, respectively, and the vertical green lines indicate their LSR velocities.}
\label{fig:CO_total}
\end{figure}
%%%%%%%%%%%%%%%%%%%%%%%%%%%%%%%%%%%%%%%%
The morphology and kinematics of the W5-NW complex is deciphered using the molecular line observations of the $J=3-2$ transitions of CO.
$^{12}$CO\,($3-2$) and $^{13}$CO\,($3-2$) spectra are extracted 
over the cloud complex 
within a circular region of radius 5.6$'$ centred at $\rm \alpha_{J2000}=02^h49^m33^s.6$; $\rm \delta_{J2000}= +60^\circ 44'43''$, (refer to green circle in {\fig}\ref{fig:CD_temp}. The averaged spectra are shown in {\fig}\ref{fig:CO_total}. Both the optically thick ($^{12}$CO\,($3-2$)) and optically thin ($^{13}$CO\,($3-2$)) transitions show two distinct emission peaks, implying two velocity components. Though the $^{13}$CO\,($3-2$) line is generally considered to be optically thin, this need not be the case in the denser regions of the cloud. To justify this assumption, we estimate the optical depth following the method outlined in \citet{2020ApJ...901...31L}. Under the assumption of a local thermodynamic equilibrium, the optical depth is calculated from the ratio of the peak intensities of $^{12}$CO\,($3-2$) and $^{13}$CO\,($3-2$) in the average spectra and the isotope ratio $^{12}$C/$^{13}$C, that is estimated to be 71 for W5-NW (see Eq. 4 of \citealt{2013A&A...554A.103P}). The optical depth estimated lies within the range 0.2 -- 0.4 for the W5-NW complex, implying the observed $^{13}$CO\,($3-2$) line transition to be optically thin across the entire region investigated here. 
Thus, the two peaks could be associated with two distinct cloud components, hereafter named as W5-NWa and W5-NWb. 
The LSR velocities from the peaks of the double Gaussian fit to the $^{13}$CO\,($3-2$) spectrum are $-41.4\,\rm km\,s^{-1}$ and $-36.3\,\rm km\,s^{-1}$ for W5-NWa and W5-NWb, respectively.
The derived line parameters (i.e., the linewidth, $\Delta V$ and velocity dispersion, $\sigma_{\rm line}$) from the Gaussian fit of these components are tabulated in {\tab}\ref{tab:CO_par}.
%%%%%%%%%%%%%%%%%%%%%%%%%%%%%%%%%%%%%%%%
\begin{table}
    \centering
    \caption{Line parameters of the $^{13}$CO\,($3-2$) and $^{12}$CO\,($3-2$) transitions towards W5-NWa and W5-NWb.}
    \begin{tabular}{c c c c c c c} \hline \hline
         Cloud & $V_{\rm LSR}$ &\multicolumn {2}{c}{$^{13}$CO\,($3-2$)}     &  \multicolumn {2}{c}{$^{12}$CO\,($3-2$)}   \\
         component  &   & $\Delta V$    & $\sigma_{\rm line}$ & $\Delta V$   & $\sigma_{\rm line}$  \\
         & (km\,s$^{-1}$)  &  (km\,s$^{-1}$)   &  (km\,s$^{-1}$)     &  (km\,s$^{-1}$)   &  (km\,s$^{-1}$) \\
         \hline \
         W5-NWa  & -41.4 & 3.1   & 1.3   & 3.9   & 1.7\\
         \hline \
         W5-NWb   & -36.3 & 3.7   & 1.6   & 4.0   & 1.7 \\
         \hline 
    \end{tabular}
    \label{tab:CO_par}
\end{table}
%%%%%%%%%%%%%%%%%%%%%%%%%%%%%%%%%%%%%%%%

The optically thick $^{12}$CO\,($3-2$) line is considered to be an excellent tracer of the spatial extent of molecular clouds. The $^{13}$CO\,($3-2$) transition, on the other hand, being optically thin, traces the relatively denser regions of the molecular cloud. We construct the integrated intensity maps of W5-NWa and W5-NWb for these transitions over the velocity ranges $-46.1$ to $-39.5\,\rm km\,s^{-1}$ and $-38.3$ to $-31.0\,\rm km\,s^{-1}$, respectively, indicated as shaded blue and red regions in {\fig}\ref{fig:CO_total}.
These regions are so chosen to effectively map W5-NWa and W5-NWb while avoiding the interaction region. The red limit for W5-NWa and the blue limit for W5-NWb are taken to be $\sim V_{\rm LSR_a}+\Delta V_{\rm a}/2$ and $\sim V_{\rm LSR_b}-\Delta V_{\rm b}/2$, respectively. 
Here $\Delta V_{\rm a}$ and $\Delta V_{\rm b}$ are the linewidths of the $^{13}$CO\,($3-2$) transition from the cloud components W5-NWa and W5-NWb, respectively, derived from the double Gaussian fit to the spectrum in \fig\ref{fig:CO_total}
and $V_{\rm LSR_a}$ and $V_{\rm LSR_b}$ are their respective LSR velocities, taken from {\tab}\ref{tab:CO_par}.
{\fig}\ref{fig:moment0} shows the integrated intensity maps for $^{12}$CO\,($3-2$) (top panel; a-c) and $^{13}$CO\,($3-2$) (bottom panel; d-f).
To validate if these velocity ranges efficiently map the distinct morphology of different cloud components, we follow the technique used by \citet{2021PASJ...73S.256E}. In this approach, new data cubes are reconstructed from the Gaussian fits and the integrated intensity maps are compared with those generated from the original data. The maps show good agreement, thus confirming that the velocity ranges used effectively sample the two clouds and the morphological features associated with them (see Appendix~\ref{appndx:vel-range-2-clouds} for details).
{\fig}\ref{fig:moment0}(a) and (c) show the integrated intensity maps sampling W5-NWa and W5-NWb, respectively, and {\fig}\ref{fig:moment0}(b) is constructed over the intermediate velocity range, $-39.5$ to $-38.3\,\rm km\,s^{-1}$. 
The line emission from W5-NWa ({\fig}\ref{fig:moment0}a) shows an overall extended morphology with a narrow U-shaped cavity opening towards the north-west direction. The emission from W5-NWb displays a filamentary structure ({\fig}\ref{fig:moment0}c). In comparison, the $^{13}$CO\,($3-2$) line mostly probes the denser regions of the cloud.

\subsection{Mass estimate of W5-NW cloud complex}
\label{sec:result2}
%%%%%%%%%%%%%%%%%%%%%%%%%%%%%%%%%%%%%%%%
\begin{table}
    \centering
    \caption{Derived physical parameters of the W5-NW cloud complex.}
    \begin{tabular}{c c c c c} \hline \hline
        $R$ & $N({\rm H_2})$ & $M$   & $M_{\rm vir}$\\
         (pc)    & (10$^{21}$\,cm$^{-2}$)   & ($10^3\,M_\odot$)     &  ($10^3\,M_\odot$)  \\
         \hline \
         3.3   & 9.0   & 6.7   & 8.1 \\
         \hline 
    \end{tabular}
    \label{tab:mass}
\end{table}
%%%%%%%%%%%%%%%%%%%%%%%%%%%%%%%%%%%%%%%%
The total mass of the W5-NW cloud complex is estimated from the hydrogen column density map that is constructed using the {\it Herschel} Hi-Gal survey images. Assuming the emission to be optically thin in the far-infrared (FIR) wavelength range of 160-500\,{\um},
we model flux densities at these wavelengths with a modified blackbody function following the formalism discussed in \citet{{2016ApJ...818...95L},{2017A&A...602A..95L}} and \citet{{2019MNRAS.485.1775I},{2020MNRAS.497.5454I}}.
Here, a pixel wise spectral energy distribution modelling is carried out keeping the dust temperature and column density, $N(\rm H_2)$ as free parameters. The opacity is taken to be $\kappa_\nu = 0.1(\nu/1200\,\rm GHz)^\beta\,cm^2\,g^{-1}$ \citep{1983QJRAS..24..267H}, where $\beta$ is the dust emissivity spectral index. We consider $\beta = 2 $ \citep{1990AJ.....99..924B}, a typical value estimated in star-forming regions.
The generated column density map is shown in {\fig}\ref{fig:CD_temp}.

The mass of the cloud complex is estimated by considering the area within the 3$\sigma$ ($\rm \sigma = 5.73\,K\,km\,s^{-1}$) contour of the $^{12}$CO\,($3-2$) integrated intensity map over the velocity range $-46.1$ to $-31.0\,\rm km\,s^{-1}$. From the estimated mean line-of-sight $\rm H_2$ column density of $9.0 \times 10^{21}\,\rm cm^{-2}$, the total mass of the cloud complex is computed using the following expression
\begin{equation}
    M=N({\rm H_2})\, \mu_{\rm H_2}\, A\, m_{\rm H},
\end{equation}
where $N({\rm H_2})$ is the mean column density, $A$ is the physical area of the cloud complex and $m_{\rm H}$ is the mass of atomic hydrogen. Taking the mean molecular weight, $\mu_{\rm H_2}$ to be 2.8, the total mass of the cloud is calculated to be $6.7 \times 10^3\,M_\odot$.

To get a better understanding of the physical condition of the W5-NW complex, we proceed to investigate the gravitational stability of the cloud complex.
For this, we compare the total mass of the cloud with its virial mass.
The virial mass of the cloud complex is estimated using the expression \citep{2011A&A...530A.118P}
\begin{equation}
    M_{\rm vir} = \frac{5 \sigma_{\rm line}^2 R}{G},
\end{equation}
where $G$ is the gravitational constant, $R$ ($=\sqrt{A/\pi}$) is the effective radius of the cloud, and $\sigma_{\rm line}$ ($\rm =1.5\,km\,s^{-1}$) is the mean of the velocity dispersion values given in {\tab}\ref{tab:CO_par} for the $^{13}$CO\,($3-2$) spectrum. The derived physical quantities of the W5-NW cloud complex are tabulated in {\tab}\ref{tab:mass}.
The virial mass of the cloud is computed to be $8.1 \times 10^3\,M_\odot$, which is marginally higher (by a factor of 1.2) than the total mass of the cloud. Within the uncertainties involved in the mass estimations, it is difficult to comment on whether the two clouds, W5-NWa and W5-NWb, are gravitationally bound or their physical association is a chance event. In the analysis that follows, we investigate the cloud kinematics to search for signatures of interaction, if any, between these cloud components.

\section{Discussions}
\label{sec:discussion}
\subsection{Signatures of cloud-cloud collision}
\label{sec:signatures}
In a recent review, \citet{{2021PASJ...73S...1F}} have discussed ``complementary distribution'' and ``bridges'' as bona fide observational signatures of CCC. In this section, we explore these for the W5-NW cloud complex from the CO line observations.

\subsubsection{Complementary distribution and gas kinematics}
\label{sec:morphology_complex}
%%%%%%%%%%%%%%%%%%%%%%%%%%%%%%%%%%%%%%%%
\begin{figure*}
\centering 
\includegraphics[scale=0.40]{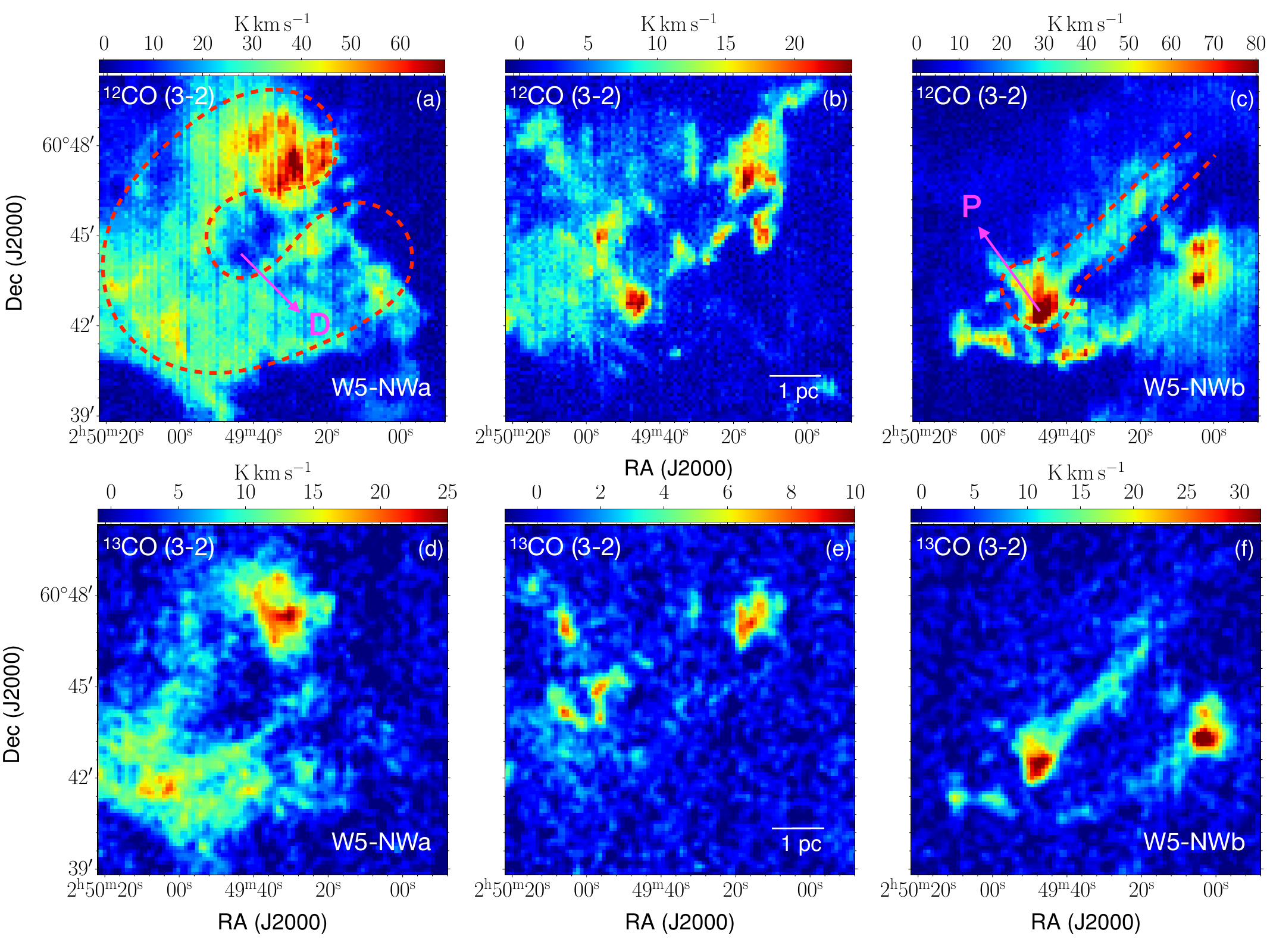}
\caption{$^{12}$CO\,($3-2$) (top panel) and $^{13}$CO\,($3-2$) (bottom panel) integrated intensity maps of the region associated with the W5-NW complex. (a) $^{12}$CO\,($3-2$) integrated intensity map over the velocity range $-46.1$ to $-39.5\,\rm km\,s^{-1}$, corresponding to W5-NWa. (b) Same as (a) for the intermediate velocity range, $-39.5$ to $-38.3\,\rm km\,s^{-1}$. (c) Same as (a) with the integration velocity range, $-38.3$ to $-31.0\,\rm km\,s^{-1}$, corresponding to W5-NWb. The depression in W5-NWa and the peak in W5-NWb are marked as D and P, respectively. (d), (e), and (f) are for $^{13}$CO\,($3-2$) in the same velocity ranges as (a), (b), and (c), respectively. The $^{13}$CO\,($3-2$) integrated intensity maps are boxcar smoothed across 3 pixels.T he schematic of the likely keyhole and key is shown in dashed lines in panels (a) and (c), respectively.}
\label{fig:moment0}
\end{figure*}
%%%%%%%%%%%%%%%%%%%%%%%%%%%%%%%%%%%%%%%%
%%%%%%%%%%%%%%%%%%%%%%%%%%%%%%%%%%%%%%%%%
\begin{figure}
    \centering
    \includegraphics[scale=0.5]{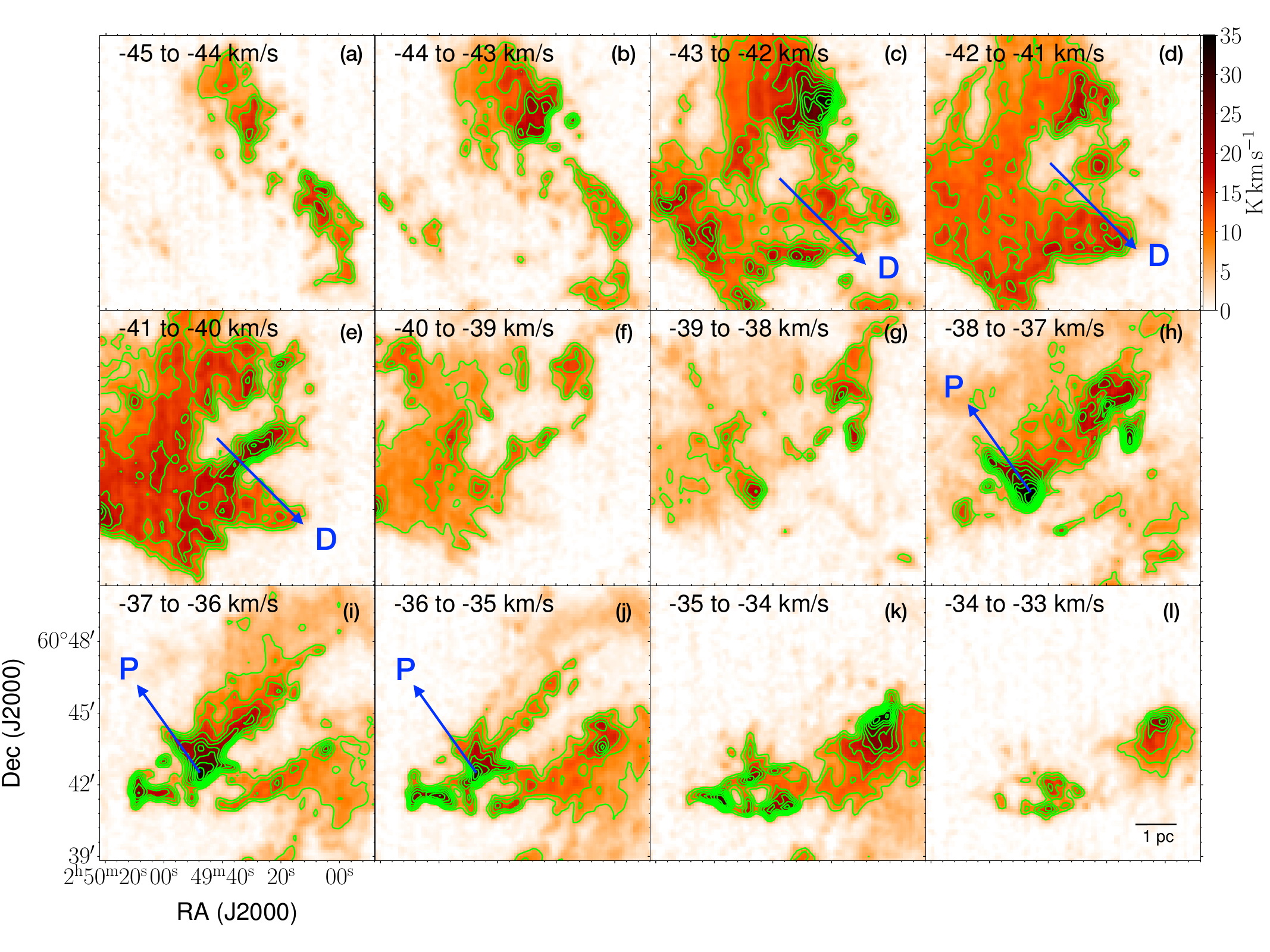}
    \caption{$^{12}$CO\,($3-2$) channel maps towards the W5-NW cloud complex. The contours start from 10$\sigma$ and increases in steps of 5$\sigma$ ($\sigma = 0.6\,\rm K\,km\,s^{-1}$). Each channel has width of $1\,\rm km\,s^{-1}$. D and P are the peak and intensity depression, respectively, of the $^{12}$CO\,($3-2$) emission. The maps are boxcar smoothed across 3 pixels.}
    \label{fig:channel_map}
\end{figure}
%%%%%%%%%%%%%%%%%%%%%%%%%%%%%%%%%%%%%%%%%%%%%
%%%%%%%%%%%%%%%%%%%%%%%%%%%%%%%%%%%%%%%%%
\begin{figure*}
\centering 
\includegraphics[scale=0.22]{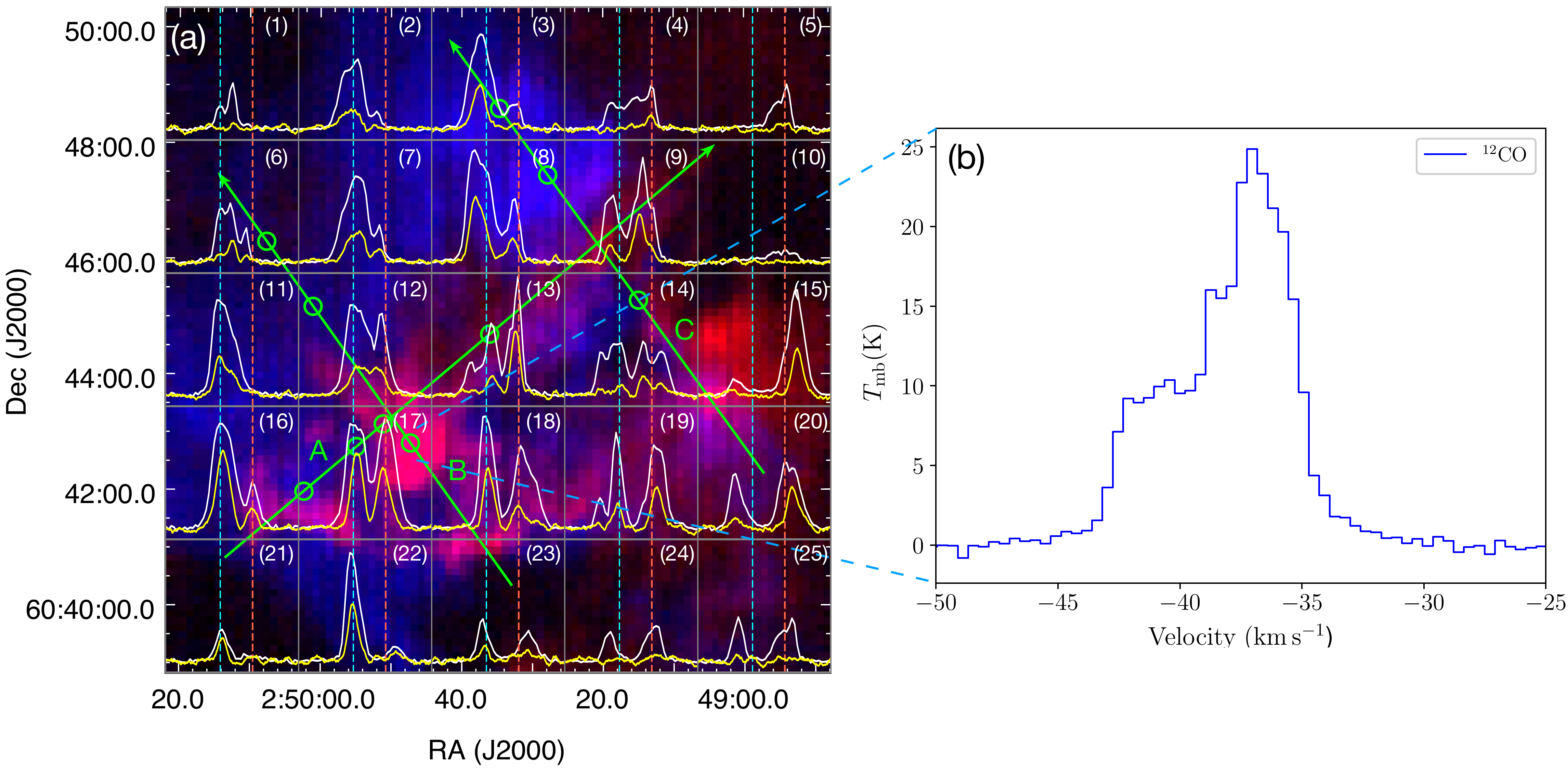} 
\caption{(a) Two-colour composite image made with the $^{12}$CO\,($3-2$) integrated intensity maps towards W5-NW. The blue and red correspond to the maps constructed within the velocity range $-46.1$ to $-39.5\,\rm km\,s^{-1}$ (W5-NWa; {\fig}\ref{fig:moment0}a) and $-38.3$ to $-31.0\,\rm km\,s^{-1}$ (W5-NWb; {\fig}\ref{fig:moment0}c), respectively. A grid of the $^{12}$CO\,($3-2$) and $^{13}$CO\,($3-2$) spectra over square regions of size $2.3'\times2.3'$ are plotted in white and yellow, respectively. The $^{13}$CO\,($3-2$) spectra are scaled-up by 1.5 times and are boxcar smoothed by 20 channels, corresponding to a velocity smoothing of $\sim 1.1\,\rm km\,s^{-1}$. The dashed blue and red lines indicate the LSR velocities of W5-NWa and W5-NWb at $-41.4\,\rm km\,s^{-1}$ and $-36.3\,\rm km\,s^{-1}$, respectively. The green arrows, labelled A, B, and C are the directions along which the PV slices are extracted, and the green circles mark the positions of the bridging features identified from the PV diagrams in {\fig}\ref{fig:pv_diagram}. (b) The $^{12}$CO\,($3-2$) spectrum extracted over a bridging feature on the PV cut B which shows a skewed single emission profile.}
\label{fig:grid_map}
\end{figure*}
%%%%%%%%%%%%%%%%%%%%%%%%%%%%%%%%%%%%%%%%
Theoretical simulations \citep[e.g.][]{{1992PASJ...44..203H},{2010MNRAS.405.1431A},{2014ApJ...792...63T},{2018PASJ...70S..58T}}
have postulated that the collision between two non-identical clouds leads to the creation of a cavity in the larger cloud by the smaller cloud. This results in a complementary distribution of two clouds at different velocities. 
As described in Fig. 3 of \citet{2018ApJ...859..166F}, based on the simulations by \citet{2014ApJ...792...63T}, once the collision occurs, a compressed layer is formed in the interacting region, creating a U-shaped cavity opening in the direction of the smaller cloud. Such a complementary distribution is accompanied by a displacement between the cavity and the smaller cloud due to projection effect, and it disappears if the projection angle is $\sim 0^\circ$ \citep[][and references therein]{2021PASJ...73S...1F}.
\citet{2018ApJ...859..166F} have identified three pairs of complementary distribution between the colliding clouds in the region of the Orion Nebula Cluster. The major complementary distribution is where the blueshifted cloud is surrounded by the U-shaped redshifted components. The other complementary distribution seen resembles a key and keyhole situation, referred to as the ``Orion key and keyhole'', wherein the blueshifted cloud component shows an intensity depression, and an emission feature corresponding to the depression is seen in the redshifted component with a displacement of 0.3\,pc. Observational evidences of complementary distribution with displacement between the colliding clouds and also the presence of U-shaped cavity have been identified in a few
%several 
other CCC candidates as well \citep[e.g.][]{{2015ApJ...806....7T},{2017ApJ...835..142T},{2021PASJ...73S..62S},{2021PASJ...73S...1F}}. 
Morphologically, W5-NW also resembles such a scenario.
In {\fig}\ref{fig:moment0}(a) and (c), we sketch visual outlines where the cloud W5-NWb displays a key-like shape and the cavity in W5-NWa has a keyhole shape, suggesting the above complementary distribution. 
The peak of the key-like structure of W5-NWb (referred as P in \fig\ref{fig:moment0}c) is complementary to the intensity depression in W5-NWa (referred as D in \fig\ref{fig:moment0}a).

In order to obtain an additional picture of the spatial distribution of the cloud complex through different velocities, we examine the $^{12}$CO\,($3-2$) channel map of the region.
\fig\ref{fig:channel_map} shows the $^{12}$CO\,($3-2$) channel map over a velocity range $-45$ to $-33\,\rm km\,s^{-1}$ with a channel width of $1\,\rm km\,s^{-1}$. The intensity peak, P and depression, D corresponding to the key structure in W5-NWb and keyhole in W5-NWa are marked in this figure.
The distribution of the $^{12}$CO\,($3-2$) emission varies significantly through the channels. 
In the velocity range of $\sim -45$ to $-40\,\rm km\,s^{-1}$ (panels a-e), the distribution is dominated by the cloud W5-NWa, wherein the intensity depression, D is marked in \fig\ref{fig:channel_map}(c), (d) and (e). And, in the range, $-38$ to $-33\,\rm km\,s^{-1}$ (panels h-l), W5-NWb dominates, wherein the peak, P is marked in \fig\ref{fig:channel_map}(h), (i) and (j). In the intermediate velocity maps (panels f and g), the U-shaped cavity resembling a keyhole and the filamentary key shaped cloud becomes evident.
This velocity segregated morphology seen in the channel maps is in good agreement with the complex gas kinematics illustrated in the grid map of the CO\,($3-2$) emission (\fig\ref{fig:grid_map}a) and the complimentary distribution.

To investigate the cloud kinematics, a two-colour composite image is made with the $^{12}$CO\,($3-2$) integrated intensity maps, illustrated in {\fig}\ref{fig:grid_map}(a), where blue and red correspond to the maps in {\fig}\ref{fig:moment0}(a) and (c), respectively.
A grid map of the $^{12}$CO\,($3-2$) and $^{13}$CO\,($3-2$) spectra are overlaid on this figure in white and yellow, respectively.
The spectra in each cell of the grid are an average over a cell size of $2.3'\times2.3'$. On careful scrutiny, the grid reveals that both the $^{12}$CO\,($3-2$) and $^{13}$CO\,($3-2$) have varying line profiles over the cloud complex. 
The spectra extracted towards the eastern (grid boxes 1, 6, 11, 16, 21, and 22) and northern (grid boxes 2, 3, 7, and 8) sides of the complex are dominated by the velocity component peaking at $-41.4\,\rm km\,s^{-1}$, corresponding to W5-NWa (blue) with some grids showing only this velocity component. In comparison, the spectra towards the western end (grid boxes 5, 10, and 15) are dominated by the emission from W5-NWb (red) at $-36.3\,\rm km\,s^{-1}$.
Moving towards the inner parts of the grid in {\fig}\ref{fig:grid_map}(a), the spectra within all the cells (grid boxes 9, 12, 13, 14, 17, 18, and 19) display prominent peaks at both velocities.
Correlating with the spatial distribution of the $^{12}$CO\,($3-2$) emission shown in colour scale, it is clearly discernible that such line profiles are seen along the interacting region. 

From the spectra presented in {\fig}\ref{fig:CO_total}, we estimate the separation, $v$ between the LSR velocities of the components, to be $\sim 5.1\,\rm km\,s^{-1}$.
This separation gives a lower limit to the relative velocity of the clouds. The actual relative velocity might be higher due to projection effects \citep{2015ApJ...807L...4F}. In a collision event, the turbulence is enhanced in the shocked layer between the clouds, irrespective of the direction of collision \citep{{2013ApJ...774L..31I},{2015ApJ...807L...4F}}. Hence, the velocity spread in the shocked layer can be taken as the relative collision velocity, $v_{\rm rel}$ \citep[][and references therein]{{2017ApJ...835L..14G},{2020MNRAS.499.3620I}}. 
A rough estimate of this relative velocity is obtained from the $^{13}$CO\,($3-2$) spectra of grids showing bridging features, as these represent interacting/collision region. A few grids (8, 12, 13, 17) are identified and the velocity span above the 2$\sigma$ ($\sigma = 0.3$\,K) level of the spectra extracted over these are calculated. This yields a range between, $6-8.5\,\rm km\,s^{-1}$ with an average value of $\sim 7\,\rm km\,s^{-1}$. Considering this average value, we estimate the projection angle, $\theta$ using the equation, $v = v_{\rm rel}\,\rm {cos}\theta$. It is found that the relative motion of the two clouds is $\sim 45^\circ$ with respect to the line-of-sight.

\subsubsection{Bridging features}
\label{sec:bridge}
%%%%%%%%%%%%%%%%%%%%%%%%%%%%%%%%%%%%%%%%%
\begin{figure}
\centering 
\includegraphics[scale=0.3]{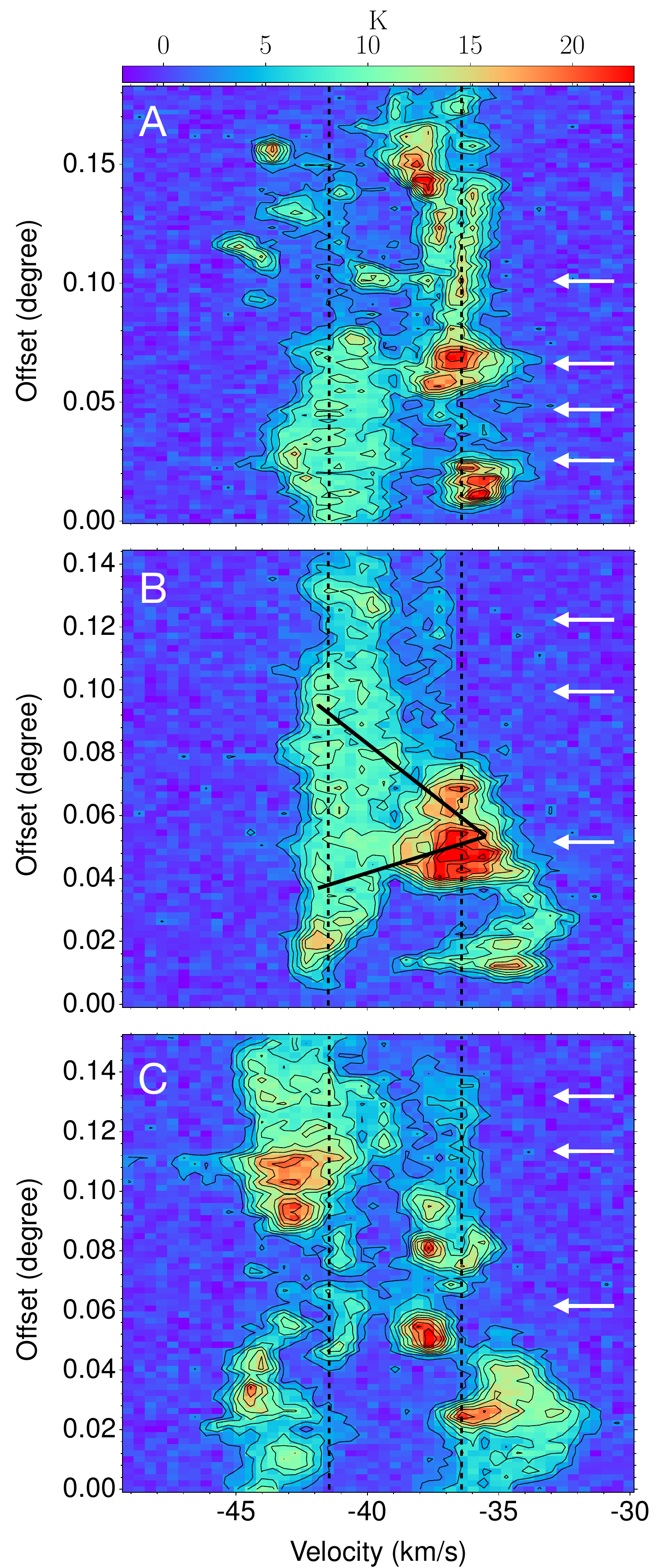}
\caption{PV diagram of $^{12}$CO\,($3-2$) along the cuts A, B, and C shown in the two-colour composite image of W5-NW in {\fig}\ref{fig:grid_map}. The contours start at 3$\sigma$ ($\sigma = 0.8$\,K) and with increments of 3$\sigma$. The identified bridging features are marked by white arrows. The V-shaped bridging feature is sketched in black along the cut B.}
\label{fig:pv_diagram}
\end{figure}
%%%%%%%%%%%%%%%%%%%%%%%%%%%%%%%%%%%%%%%%
%%%%%%%%%%%%%%%%%%%%%%%%%%%%%%%%%%%%%%%%
\begin{figure*}
\centering 
\includegraphics[scale=0.32]{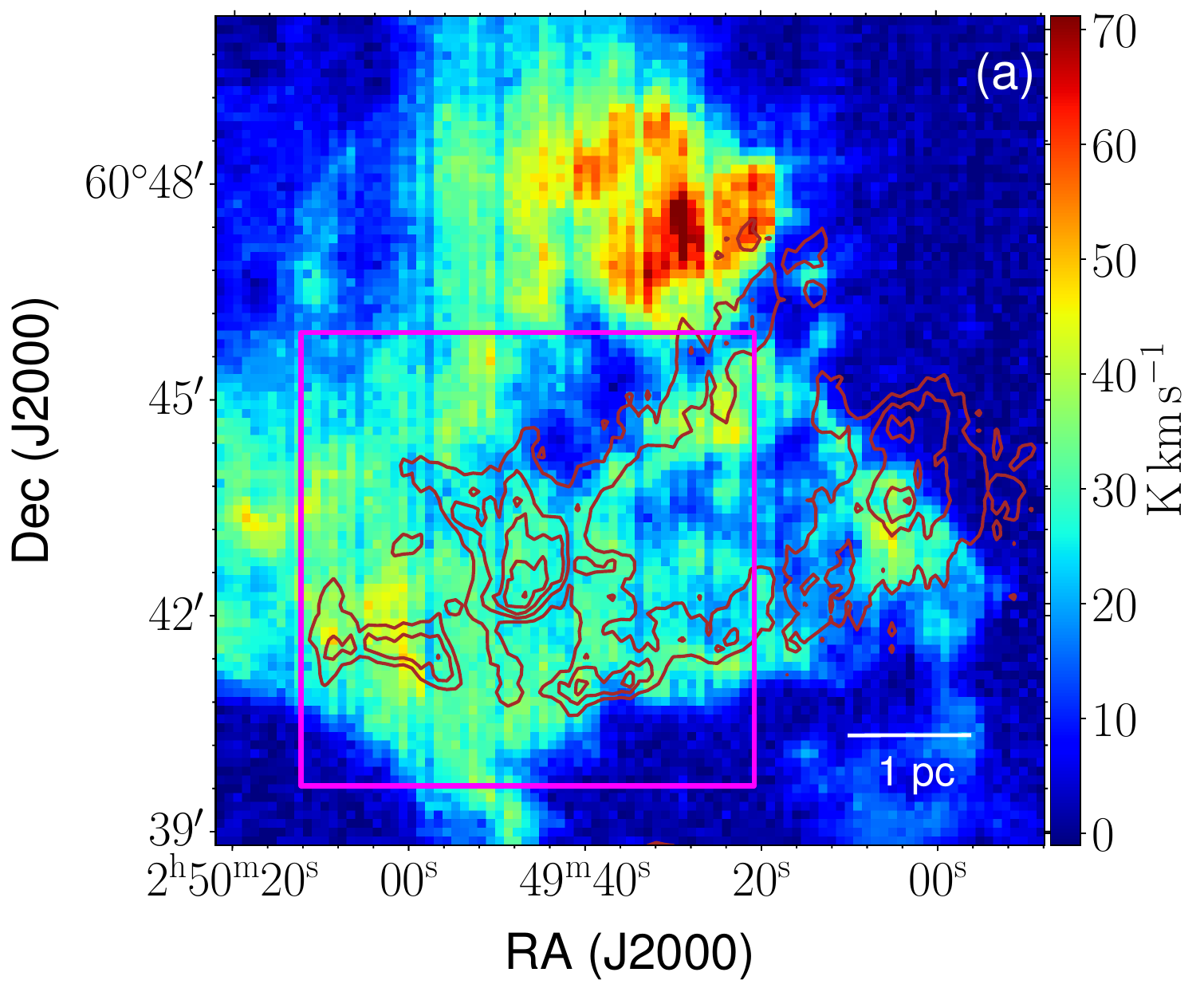} \quad\includegraphics[scale=0.32]{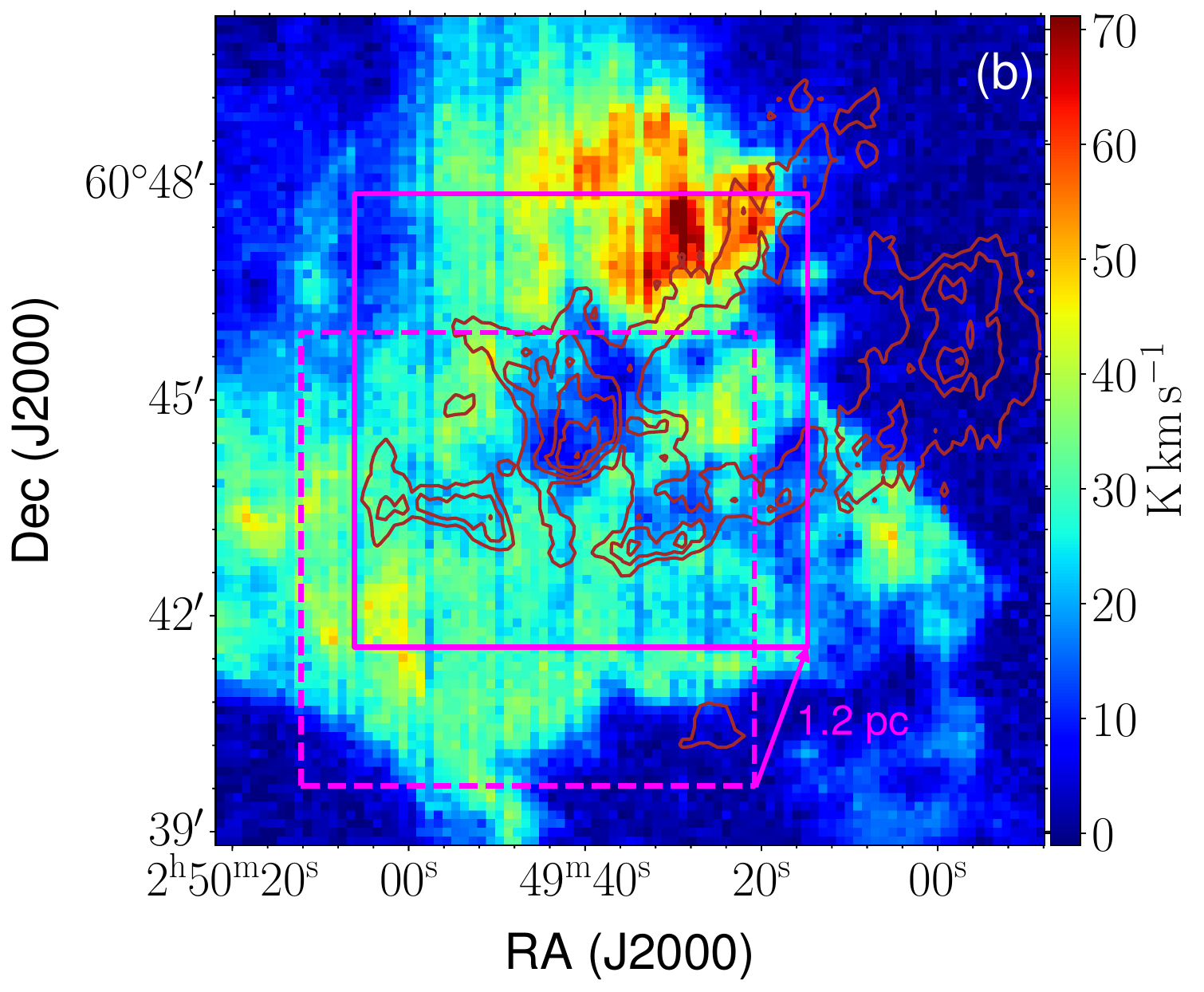}
\caption{(a) $^{12}$CO\,($3-2$) integrated intensity map of W5-NWa (colour-scale) and W5-NWb (contours). (b) Same as (a) with the contours of W5-NWb displaced by 1.2\,pc. The contours start from 10$\sigma$ and increases in steps of 7$\sigma$ ($\sigma = 2.7\,\rm K\,km\,s^{-1}$).}
\label{fig:displacement}
\end{figure*}
%%%%%%%%%%%%%%%%%%%%%%%%%%%%%%%%%%%%%%%%
Apart from the integrated intensity and channel maps, position-velocity (PV) diagrams are commonly used to describe the gas kinematics of clouds.
Typically, CCC regions are characterised by ``bridging-features'' between the two colliding clouds in the PV diagram. From the synthetic PV diagrams constructed for different astrophysical systems, including CCC models, \citet{{2015MNRAS.450...10H},{2015MNRAS.454.1634H}} have found that such bridging features are exclusive to CCCs, making them a distinct kinematic signature of colliding clouds. In the spatial distribution of the clouds, such bridging features appear at the sites of collision. Observational evidence of such bridging features can be found throughout the literature on CCC complexes \citep[e.g.][]{{2016ApJ...820...26F},{2017ApJ...835L..14G},{2020MNRAS.499.3620I},{2022ApJ...925...60D}}.

{\fig}\ref{fig:pv_diagram} displays the PV diagrams constructed towards W5-NW along three directions that samples the cloud complex. %including the interacting region.
These three directions are represented by green arrows labelled A, B, and C in {\fig}\ref{fig:grid_map}(a). 
The PV diagrams clearly show the well separated clouds W5-NWa and W5-NWb at velocities, $-41.4\,\rm km\,s^{-1}$ and $-36.3\,\rm km\,s^{-1}$, respectively. The two clouds are connected in the velocity space by ``bridging features'' with intermediate velocities, the locations of which are indicated by arrows in  {\fig}\ref{fig:pv_diagram}. The positions of these bridges are also marked on the grid map in {\fig}\ref{fig:grid_map}(a) by green circles. Such bridging features arise from the turbulent gas in the interacting region where there is an exchange of momentum between the two colliding clouds, leading to the formation of an intermediate velocity component \citep{2021PASJ...73S...1F}. Hence, at these regions, the initial velocity components of the two colliding components might not be clearly discernible, and often results in a single velocity component. Depending on the inclination angle, this velocity component can appear skewed (e.g., Fig. 2j of \citealt{2021PASJ...73S...1F}). Such a profile also manifests as a skewed ``V-shaped'' bridging feature in the PV diagram (e.g., Fig. 14 of \citealt{2018ApJ...859..166F}, Fig. 2k of  \citealt{2021PASJ...73S...1F}, and Fig. 5 of \citealt{2021PASJ...73S..62S}). A similar feature is seen in the PV diagram along the cut B in W5-NW where a clear V-shaped gas distribution is revealed, the outline of which is traced in {\fig}\ref{fig:pv_diagram}. The extracted $^{12}$CO\,($3-2$) spectrum towards this bridge, plotted in {\fig}\ref{fig:grid_map}(b), also displays the skewed single emission profile discussed above. These kinematic features identified in the PV diagrams strongly indicate a cloud collision scenario in W5-NW. The velocity structures decoded from the CO transition provide strong indication of CCC in the W5-NW complex. In recent literature \citep[e.g.][]{{2013ApJ...775...88N},{2016A&A...595A.122L}}, the low-velocity component of SiO emission has been shown to be an excellent tracer of colliding flows and CCC. Future high-resolution and deep SiO observations focused towards the identified bridging features and YSO clusters can help better understand the mechanisms involved.

From Sections~\ref{sec:morphology_complex} and \ref{sec:bridge} we find robust evidences of complementary distribution and bridging features for CCC scenario in the W5-NW complex. As postulated by \citet[][and references therein]{2021PASJ...73S...1F}, the complementary distribution of the intensity peak of the small cloud is displaced from the depression in the larger cloud for projection angles $\neq 0^\circ$. The peak, P of the key-like structure of W5-NWb shows a displacement in the south-east direction from the intensity depression, D in W5-NWa.
We estimate the displacement between P and D following the method described in \citet{2018ApJ...859..166F} and it is found that P fits within D after displacing it by $\sim 1.2$\,pc in the north-west direction at an angle of $21^\circ$ counter-clockwise from the RA axis. \fig\ref{fig:displacement} shows the $^{12}$CO\,($3-2$) integrated intensity map of W5-NWa in colour-scale and of W5-NWb as the contours. The complementary distribution of the W5-NWb key and W5-NWa keyhole can be clearly seen in \fig\ref{fig:displacement}(b) when a displacement of 1.2\,pc is applied.

\subsection{Collision induced cluster formation}
\label{sec:cluster formation}
Clustered and triggered star formation has been extensively studied in the W5 star-forming complex \citep{{2008ApJ...688.1142K},{2012A&A...546A..74D}}. This complex comprises of two distinct {\it Spitzer} bubbles around the H\,{\small II} regions W5-E and W5-W that are excited by O-type stars, HD 18326 (O7-V star) and HD 237019 (O8-V star), respectively. Based on {\it Spitzer} mid-infrared observations, \citet{2008ApJ...688.1142K} propose ongoing star formation in the W5-NW and W5-NE clouds to be sequentially triggered by an initial episode of s1 (Copy)tar formation involving HD 237019 and expansion of the H\,{\small II} region. In a later study, \citet{2012A&A...546A..74D} used {\it Herschel} far-infrared maps to arrive at a similar inference of triggered star formation in W5-E. Their analysis suggests the new generation of stars in W5-NE to have formed from dense condensations compressed between the expanding H\,{\small II} regions.
%%%%%%%%%%%%%%%%%%%%%%%%%%%%%%%%%%%%%%%%%
\begin{figure*}
\centering 
\includegraphics[scale=0.17]{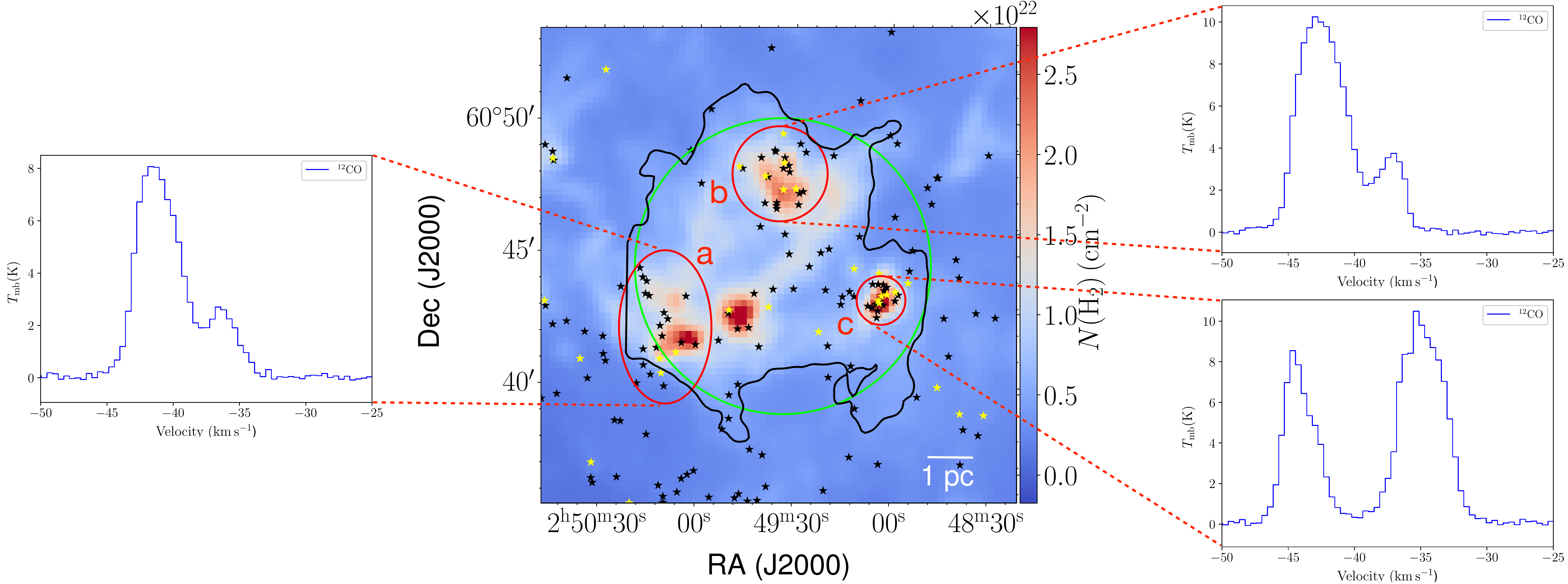}
\caption{Line-of-sight H$_2$ column density map of the W5-NW region, generated using the Hi-Gal maps from 160-500\,{\um}. This map has a resolution of 36$''$ and pixel size of 14$''$. The 3$\sigma$ ($\rm \sigma = 5.73\,K\,km\,s^{-1}$) contour of the $^{12}$CO\,($3-2$) integrated intensity map is overlaid in black. 
The green circle is the region over which the overall spectra described in {\fig}\ref{fig:CO_total} is extracted.
The yellow and black stars indicate the position of the Class I and II YSOs, respectively, identified by \citet{2008ApJ...688.1142K}. The YSO clustering regions (red ellipses) are labelled a, b, and c and the corresponding $^{12}$CO\,($3-2$) spectra extracted over these regions are shown.}
\label{fig:CD_temp}
\end{figure*}
%%%%%%%%%%%%%%%%%%%%%%%%%%%%%%%%%%%%%%%%

Evidence of CCC prompts us to explore the possibility of collision induced star formation in W5-NW.
CCC events are known to trigger star formation in the interacting regions as well as within the colliding clouds. High-mass star and super-star cluster formation have been reported in literature over recent years \citep[e.g.][and references therein]{{2020MNRAS.499.3620I},{2021PASJ...73S...1F},{2022A&A...663A..97M}}.
\citet{2008ApJ...688.1142K} investigated the evolutionary phases of identified point sources based on infrared colors and magnitudes and classified them as Class~I/Class~II/Class~III/transition disks/embedded protostars. 
From the above catalog of protostars, 34 Class~I and 203 Class~II sources lie within the field of view of the {\it JCMT} ${}^{12}$CO (3-2) map ($\sim 14'\times 14'$) covering the W5-NW complex. 

The distribution of identified YSOs is shown in {\fig}\ref{fig:CD_temp}. It is seen from the figure that within the W5-NW complex, the Class~I and II YSOs are not randomly located but clustered towards dense clumps of high column density
(marked with red ellipses) and along intra-clump filamentary features. 
Three regions of YSO clustering are evident, and these regions are labelled as
%in the dense clumps are identified as 
\textit{a}, \textit{b}, and \textit{c}.
An over-density of Class~I sources is seen towards the regions \textit{b} and \textit{c} located towards the western part of the complex, indicating that these may be young candidate clusters.
The $^{12}$CO\,($3-2$) spectra extracted towards these regions (see {\fig}\ref{fig:CD_temp}) show the presence of two velocity components, suggesting these to be located in interacting regions.

%%%%%%%%%%%%%%%%%%%%%%%%%%%%%%%%%%%%%%%%%
\begin{figure*}
\centering 
\includegraphics[scale=0.7]{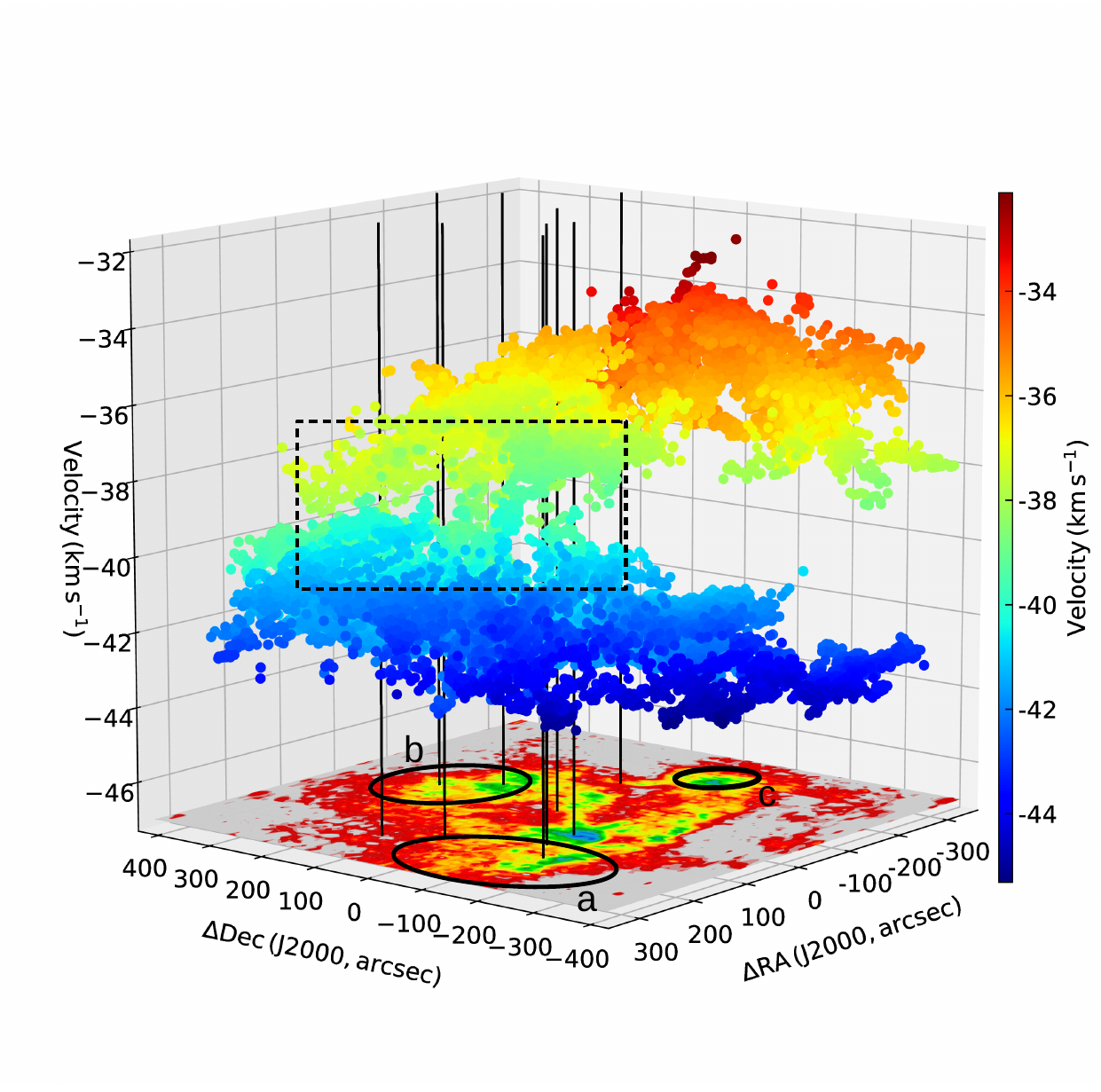}
\caption{PPV diagram of all fitted components in W5-NW. Each voxel denotes the spatial location and best-fit centroid velocity of the Gaussian component identified in $^{12}$CO (3-2) emission using SCOUSEPY. Colour of each point represents these best-fit velocities. The integrated intensity map of $^{12}$CO (3-2) integrated over the full velocity range is shown at the base of the plot. The coordinates axes are specified as relative offset from $\rm \alpha_{J2000}=02^h49^m34^s.9$; $\rm \delta_{J2000}= +60^\circ 44'42''$. The vertical lines denote the position of the green circular apertures shown in Fig \ref{fig:grid_map}. Connecting velocity structures are highlighted within the drawn rectangle. Ellipses showing YSO clusters (see {\fig}\ref{fig:CD_temp}) are marked.}
\label{fig:3d_scouse}
\end{figure*}
%%%%%%%%%%%%%%%%%%%%%%%%%%%%%%%%%%%%%%%%
To obtain a more clear picture of the identified clusters and the gas kinematics therein, we implement the Semi-automated multi-COmponent Universal Spectral-line fitting Engine \citep[SCOUSE;][]{2016MNRAS.457.2675H} algorithm, which provides an excellent tool to obtain a 3D view of the molecular gas distribution and kinematics. A python implementation of the same, SCOUSEPY \citep{2019MNRAS.485.2457H}, is used here. A brief description of the algorithm is given in the Appendix. 
The generated position-position-velocity (PPV) diagram showing the distribution of ${}^{12}$CO (3-2) gas associated with the W5-NW cloud complex is presented in {\fig}\ref{fig:3d_scouse}. 
Each data point in this 3D plot (termed as voxel) represents the position (RA, DEC) and velocity coordinates extracted by the SCOUSEPY algorithm. The vertical lines mark the location of the bridging features that are identified in the PV diagram ({\fig}\ref{fig:pv_diagram}). 
The centroid velocities of the fitted Gaussian profiles are color encoded which clearly demonstrate the presence of two distinct velocity components, W5-NWa ($-46.1$ to $-39.5\,\rm km\,s^{-1}$) and W5-NWb ($-38.3$ to $-31.0\,\rm km\,s^{-1}$) connected with the intermediate velocity features ($-39.5$ to $-38.3\,\rm km\,s^{-1}$).
The interacting regions with connecting velocity structures are highlighted within the rectangle. These mostly coincide with the vertical lines that trace the bridging features identified in the PV diagrams. The figure also shows the integrated intensity map of $^{12}$CO\,($3-2$) and the locations of the identified clusters. The clusters \textit{a} and \textit{b} are in regions displaying velocity connecting features. This is a signature of collision induced cluster formation. Though the cluster \textit{c} is not co-spatial with any bridging feature, its formation could still be triggered by a CCC event, since bridging features tend to disappear within a short timescale of the order of $10^5$\,yr after collision \citep{2015MNRAS.454.1634H}.

Further, to lend support to the above conjecture, we estimate the collision time-scale between W5-NWa and W5-NWb from the size of the cloud, 6.5\,pc, and the relative collision velocity of the two clouds, $\sim 7\,\rm km\,s^{-1}$. The ratio between the two gives a collision time-scale of $\sim 0.9$\,Myr. It should be noted that this estimate gives an order-of-magnitude at best, and the actual time-scale might vary by a factor of $\sim 2$ owing to the projection effects in space and velocity and also the unknown configuration of the clouds prior to collision \citep{2014ApJ...780...36F}. 
Typical ages of Class~I and II YSOs are of the order of 0.4-0.7\,Myr \citep{2015ApJS..220...11D} and 2$\pm$1\,Myr \citep{2009ApJS..181..321E}, respectively. Within uncertainties, the collision timescale is comparable to these. 
Observational evidence of CCC, location of YSOs, and comparable timescale of collision supports collision induced formation of the cluster of YSOs. However, one cannot rule out the role of the expanding H\,{\small II} region in triggering the collision event itself.

\section{Conclusions}
\label{sec:conclusion}
The structural and kinematic study of the star-forming complex W5-NW unveils a likely scenario of cloud-cloud collision that has in-turn triggered star formation at the collision zone, as well as a possible cluster formation. Several observational signatures of CCC are explored and are summarized below. 

\begin{enumerate}
    \item The physical association of W5-NWa and  W5-NWb is attributed to a collision event. Supporting this claim, we see complementary distribution of the two clouds, where W5-NWb has a filamentary key-like structure that fits into a U-shaped cavity in W5-NWa that acts as the keyhole with a displacement of 1.2\,pc at a position angle of $21^\circ$ from the RA axis.
    \item Several bridging features are also seen between the two clouds in the PV diagram. These features that arise from the compressed layer formed due to collision have intermediate velocities. A skewed V-shaped bridging feature is also detected at the site of collision, where the $^{12}$CO\,($3-2$) spectrum extracted has a skewed line profile as well. 
    \item The manifestation of the cloud-cloud collision is seen in the PPV diagram obtained using SCOUSEPY. 
    The connecting velocity structures in the three-dimensional image provide a robust signature of the interaction.
    \item The distribution of identified Class~I and II YSOs in the interacting region gives compelling evidence of collision induced cluster formation in W5-NW. This is supported by the collision timescale of $\sim 0.9$\,Myr, which is comparable to the typical ages of the Class~I and II YSOs.
\end{enumerate}

The investigation of gas kinematics in the W5-NW cloud emphasizes the need for such observational studies to firmly establish signatures of CCC and complement the theoretical framework of collision, mass compression, and star/cluster.

%% Also note that the akcnowlodgment environment does not support long amounts of text. If you have a lot of people and institutions to acknowledge, do not use this command. Instead, create a new \section{Acknowledgments}.
\begin{acknowledgments}
We would like to thank the referee for valuable comments/suggestions. This work has been supported by the National Key R\&D Program of China (No. 2022YFA1603100). H.-L. Liu is supported by National Natural Science Foundation of China (NSFC) through the grant No.12103045, and by Yunnan Fundamental Research Project (grant No.\,202301AT070118). T. Liu acknowledges the supports by NSFC through grants No.12073061 and No.12122307, the international partnership program of Chinese academy of sciences through grant No.114231KYSB20200009, and Shanghai Pujiang Program 20PJ1415500. This paper makes use of the data obtained with {\it JCMT}. The {\it JCMT} is operated by the East Asian Observatory on behalf of the National Astronomical Observatory of Japan, Academia Sinica Institute of Astronomy and Astrophysics, the Korea Astronomy and Space Science Institute and Center for Astronomical Mega-Science (as well as the National Key Research and Development Program of China with No. 2017YFA0402700). Additional funding support is provided by the Science and Technology Facilities Council of the United Kingdom and participating universities in the United Kingdom and Canada. This publication also makes use of data products from Herschel (ESA space observatory).
\end{acknowledgments}

\appendix
\restartappendixnumbering

\section{Representative velocity-range for the two clouds}
\label{appndx:vel-range-2-clouds}
In clouds with a complex velocity structure containing multiple components, it is challenging to distinguish these components due to the close proximity of their velocities. This %challenge 
is evident in the grid map shown in Fig. \ref{fig:grid_map}, where the separation between the velocity peaks of the red and blue cloud varies significantly across each grid. Consequently, any two fixed velocity ranges alone cannot accurately depict the distribution of molecular emission.

\begin{figure*}
\centering 
\includegraphics[scale=0.7]{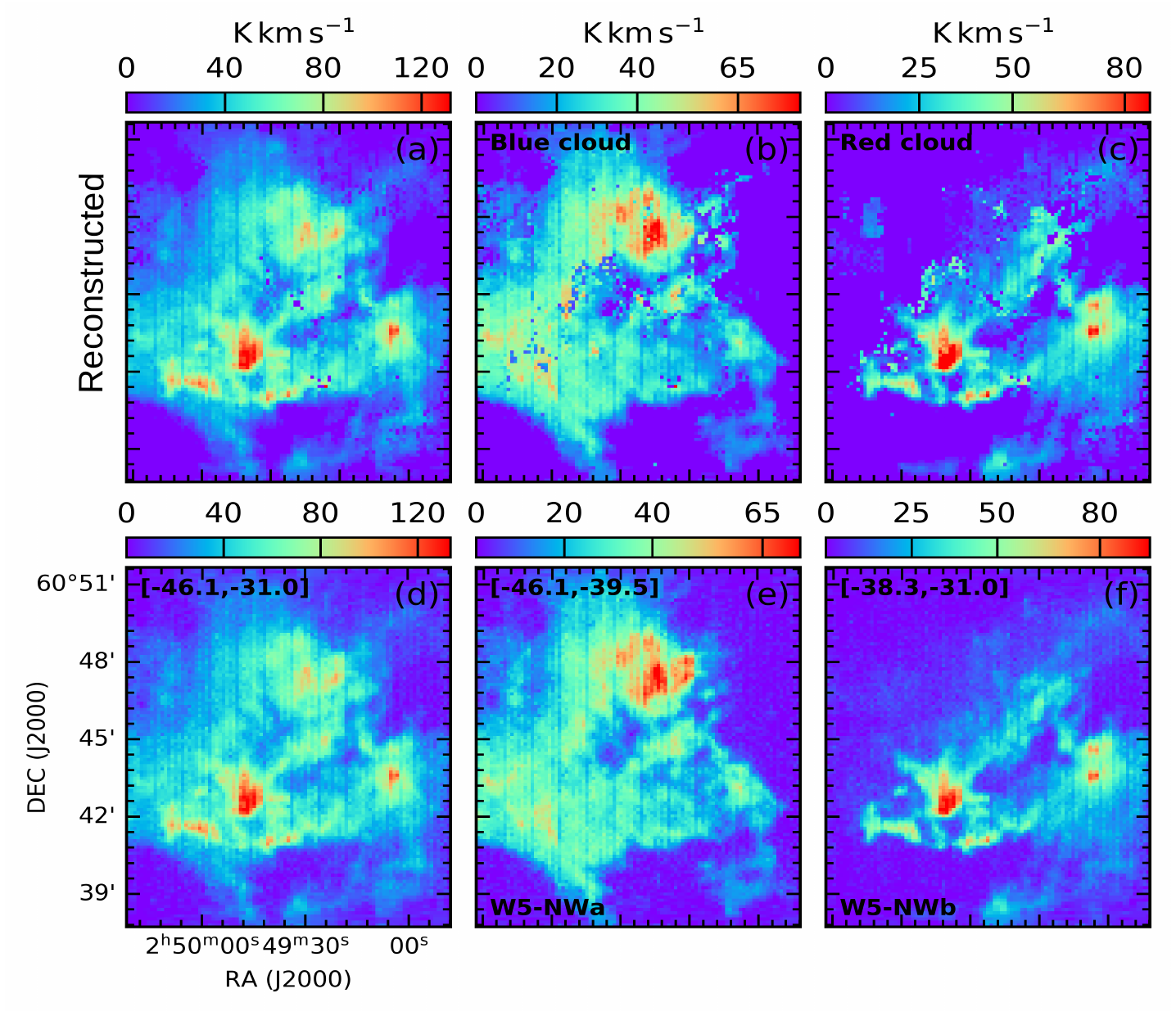}
\caption{Integrated intensity distribution of the original $^{12}$CO\,($3-2$) cube and the reconstructed cube. (a), (b) and (c) Integrated intensity distribution of the full reconstructed cube, reconstructed blue cloud, and reconstructed red cloud, respectively. (d), (e) and (f) Integrated intensity map of original $^{12}$CO\,($3-2$) cube over velocity ranges corresponding to W5-NW complex, W5-NWa and W5-NWb, respectively. The velocity ranges used to obtain the maps are mentioned in top left of panels (d) - (f). Panels (e) and (f) are same as Fig. \ref{fig:moment0} (a) and (c).
Pixels where Gaussian fitting is not possible are not considered in the analysis and are excluded in this plot.}
% \textbf{A few pixels where we are unable to fit Gaussian functions are not considered in our analysis.}}
\label{fig:reconstructed-mom-maps}
\end{figure*}

To address this, we followed the method outlined in Appendix 2 of \citet{2021PASJ...73S..75E}. To implement this, we have used the SCOUSEPY algorithm discussed in Section \ref{sect:scouse} to fit Gaussian functions to the spectrum from each pixel to establish the representative velocity ranges for the two clouds and reconstructed a new data cube consisting of only the fitted Gaussian profiles. In \fig\ref{fig:reconstructed-mom-maps}, we present the integrated intensity map from the original $^{12}$CO\,($3-2$) cube and from the reconstructed cube. The reconstructed data effectively replicates the morphology of the W5-NW complex. Furthermore, to see the spatial distribution of the blue and the red cloud in the reconstructed cube, we grouped the fitted profiles separately into two single Gaussian profiles.
The reconstructed blue cloud consists of (i) the lower-velocity profile if the fitted spectrum for a pixel has double Gaussian profile, and (ii) if the fitted profile is single Gaussian, the profile with the centroid velocity less than -38.5 $\rm km\,s^{-1}$. Conversely, the reconstructed red cloud includes (i) the higher-velocity profile if the fitted spectrum for a pixel has double Gaussian profile, and (ii) if the fitted profile is single Gaussian, the profile with the centroid velocity greater than -38.5 $\rm km\,s^{-1}$. 
The integrated intensity distribution of the blue and red clouds shown in \fig\ref{fig:reconstructed-mom-maps}(b) and (c), respectively, resemble W5-NWa (\fig\ref{fig:reconstructed-mom-maps}e) and W5-NWb (\fig\ref{fig:reconstructed-mom-maps}f).
This comparison suggests that the velocity ranges ($-46.1$ to $-39.5\,\rm km\,s^{-1}$ for W5-NWa and $-38.3$ to $-31.0\,\rm km\,s^{-1}$ for W5-NWb) selected for both the cloud components effectively represent all distinct morphological features within the clouds.
%%%%%%%%%%%%%%%%%%%%%%%%%%%%%%%%%%%%%%%%%%%
%%%%%%%%%%%%%%%%%%%%%%%%%%%%%%%%%%%%%%%%%%%

\section{Brief outline of the SCOUSEPY algorithm}
\label{sect:scouse}
SCOUSEPY works in several steps to fit complex spectroscopic data efficiently and systematically. A detailed description of this algorithm is given in \citet{2016MNRAS.457.2675H}, here we briefly outline the same. Firstly, this algorithm identifies regions over which the fitting is carried out. A specific region of the data having values above a given noise level can be selected in both position and velocity axes. Subsequently, SCOUSEPY splits that significant region into smaller areas called spectral averaging areas (SAAs) based on the input parameter ‘wsaa', which is the width of the SAA. Spatially averaged spectra for each SAA are then fitted interactively using PYSPECKIT to get the best-fit parameters. Spectrum extracted over each pixel inside the SAAs is then fitted automatically using the best-fit parameters from the manual fit of their SAAs within specified tolerance levels. Tolerance levels limit the peak line intensity of all detected components, FWHM, centroid velocity of the line, and separation between the identified components. 
The next step involves checking the best-fit spectrum of each pixel, using diagnostic plots of \textit{rms}, residuals, reduced chi-square, and number of components.
In the final step, anomalous spectra can be re-fitted manually. This multi-Gaussian fitting approach provides information on peak intensity, central velocity, and linewidth of all velocity components in all emission lines. 

For the analysis of the W5-NW complex, we consider a square area of side $\sim$42 arcsec for each SAA. It represents the spatial area over which the algorithm extracts the averaged spectrum. Each SAA contains 36 spectra. Reducing the size of SAA does not significantly impact the final best-fitted parameters but increases the number of spectra to be fitted manually in the first stage.
The tolerance levels are summarised as follows: (i) peak intensity of each detected component must be greater than three times the \textit{rms}; (ii) minimum FWHM of each Gaussian component must be two channel spacing; (iii) velocity and line width parameters derived in the automated fitting stage must be within a factor of three of the corresponding values derived in the interactive fittings of each SAA; (iv) velocity components, separated by at least the FWHM of the narrowest component are considered to be distinct. 

%%%%%%%%%%%%%%%%%%%%%%%%%%%%%%%%%%%%%%%%%%%%%%%%%%


\bibliography{reference}{}
\bibliographystyle{aasjournal}



\end{document}